\documentclass[prb,twocolumn]{revtex4}

\usepackage{graphicx}
\usepackage{dcolumn}
\usepackage{amsmath}
\def\Khg{$\alpha$-(BEDT-TTF)$_2$KHg(SCN)$_4$}
\def\DelT{$\mathit{\nabla T}$}
\newcommand{\tep}[2][]{$S_{#1}^{#2}$}
\makeatother
\begin{document}
\title{Magnetothemopower study of quasi two-dimensional
organic conductor $\alpha$-(BEDT-TTF)$_2$KHg(SCN)$_4$}
\author{E. S. Choi}
\affiliation{Department of Physics, Ewha Womans University, Seoul
120-750, Korea and National High Magnetic Field Laboratory,
Florida State University, Tallahassee, Florida 32310}
\author{J. S. Brooks}
\author {J. S. Qualls}
\thanks{Present address: Physics Dept., Wake Forest University,
Winston-Salem, NC
  27109}
\affiliation{National High Magnetic Field Laboratory and Physics
Department, Florida State University, Tallahassee, Florida 32310 }
\date{\today}
\begin{abstract}
We have used a low-frequency magneto-thermopower (MTEP) method to
probe the high magnetic field ground state behavior of
$\alpha$-(BEDT-TTF)$_2$KHg(SCN)$_4$ along all three principal
crystallographic axes at low temperatures. The thermopower tensor
coefficients ($S_{xx}, S_{yx}$ and $S_{zz}$) have been measured to
30 T, beyond the anomalous low temperature, field-induced
transition at 22.5 T. We find a significant anisotropy in the MTEP
signal, and also observe large quantum oscillations associated
with the de Haas - van Alphen effect. The anisotropy indicates
that the ground state properties are clearly driven by mechanisms
that occur along specific directions for the in-plane electronic
structure. Both transverse and longitudinal magnetothermopower
show asymptotic behavior in field, which can be explained in terms
of magnetic breakdown of compensated closed orbits.
\end{abstract}

\maketitle
\section{Introduction}
The ground state of the quasi-two dimensional organic conductor
\Khg~ has emerged as one of the more important fundamental
problems in the area of synthetic metals. This is due to the
highly unusual magnetic field dependent behavior of the low
temperature ground state which appears below a transition
temperature $T_p$ = 8 K. The band calculation and the Fermi
surface study\cite{mori1} suggest a two dimensional Fermi surface
consisting of a quasi-one dimensional electron section and a
closed hole pocket as shown in Fig. \ref{fig1} (a). The density
wave formation is believed to originate from the instability of
the open orbit at low temperatures, which is followed by Fermi
surface reconstruction\cite{kartsovnik}. This in turn leads to the
unusual behavior of this compound at low temperatures and high
magnetic field\cite{kartsovnik2,sasaki1}. This ground state is now
thought to be related to a charge density wave state where both
spin and orbital mechanisms compete, as evidenced by recent
studies in tilted magnetic fields\cite{qualls,proust}. Above 22.5
T an anomaly (called the "kink field" or $B_k$ after its discovery
by Osada et al., \cite{osada} which appears in various
transport-related measurements) occurs, and a new high field
electronic state emerges. Recently the observation of critical
state-like behavior in the magnetization\cite{harrison1} and
susceptibility \cite{harrison2} above $B_k$ has been reported,
which indicates an unusual, highly conductive state which is
associated with the Landau level filling fraction. Additional
thermodynamic evidence for unusual, hysteretic behavior above
$B_k$, which is related to the Landau level filling, has also been
reported in high field magnetocaloric measurements \cite{fortune}.

The purpose of the present work has been to probe this novel
ground state by thermoelectric power (TEP), which involves
measuring the potential difference across a sample in the presence
of a thermal gradient, as shown in Fig. 1 (c). In general, the
resulting signal (in V/K) has the form\cite{mcdonald}
\begin{equation}
\label{mott} S = \frac{\pi^2 k_B^2}{3e}T (\frac{d \mathrm{ln}
n(E)}{dE}+ \frac{d \mathrm{ln} v^2(E)}{dE}+\frac{d
\mathrm{ln}\tau(E)}{dE})\arrowvert_{E= E_{\rm F}}
\end{equation}
where $n(E)$ is the density of the states, $v(E)$ is an average
charge velocity, and $\tau(E)$ is the carrier scattering
relaxation time. In addition, we have also measured the Nernst
effect for specific sample configurations. This involves the
transverse thermopower which is given in the simple approximation
by $S_{xy} = Q B (l/d)$, where $Q$ is the Nernst coefficient, $B$ is
the magnetic field, and $l/d$ is the ratio of the distance between
the Nernst leads and the sample thickness. We note that ideally, for closed orbits,
$S_{xy}$ is not sensitive to electron/hole compensation effects,
and that it increases linearly with magnetic field. (See
discussion below).

Clearly the TEP and Nernst signals can be a complex mixture of
phenomena which can be difficult to interpret, unless specific
aspects of the system under consideration dominate and/or change
dramatically\cite{chaikin}. Examples include the case of a
conventional metal ($S \approx k_B/e ~k_B T /E_{\rm{F}}$), a
superconductor ($S = 0$), and the quantum Hall effect ($S \approx$
0 or a finite value depending on the Landau level filling). As we
will show in the present case, the TEP probe for \Khg~ is
particularly sensitive to the Landau level spectrum, to the gap in
the electronic structure, and to the in-plane electronic
anisotropy.

In light of the discussion above, we state here the main findings
of the present work before the full presentation.
\underbar{First}, where comparisons can be made with previous
studies (all of which are in-plane, and almost exclusively at low
or zero magnetic fields), we find that our results agree in the
details of the sensitivity of the thermopower to the anisotropic
Fermi surface topology, and with absolute value of the thermopower
signals previously reported. In addition, we extend these
measurements below $T_p$, where the TEP reveals details of the
transition into the density wave state. \underbar{Second}, we
observe, in the oscillatory METP, quantum oscillations associated
with the main closed Fermi surface orbit frequency, and also
oscillations associated with Fermi surface reconstruction and
anomalous second harmonic signals which result from the density
wave ground state. We believe that this is the first detailed
study of quantum oscillations in a quasi-two dimensional organic
conductor by thermopower methods.  \underbar{Third}, the
background (non-oscillatory) METP is sensitive to changes in the
density wave ground state with magnetic field. This includes the
signatures for three changes in the Fermi surface topology, one near
3.75 T, a second above 10 T, and a third above 22.5 T. The first
two are associated with magnetic breakdown effects, and the latter
is the first order phase transition to the high field state. The
temperature-magnetic field phase diagram, based on the METP
results, is presented. \underbar{Fourth}, we study in detail for
the first time the inter-plane thermopower, which gives additional
information about the role of the inter-plane transport mechanism
of this material. \underbar{Finally}, in the high field state, we
describe the phase relationship between the Nernst signal and the METP signal,
which is consistent with theoretical predictions.

\section{Background}
To our knowledge, no thermopower measurements have been carried
out on the \Khg ~ material in very high magnetic fields. In
1950's, a series of calculations were done for high field
dependence of magnetotransport properties. It was found that, in
the limit of $\omega_c\tau \gg 1$ (where $\omega_c$ is the
cyclotron frequency and $\tau$ is the scattering time), asymptotic
magnetotranport properties (electrical conductivity\cite{LAK},
thermal conductivity\cite{azbel} and thermopower\cite{BGN}) are
highly dependent on the Fermi surface topology and carrier
compensation, while they weakly depend on the scattering time and
the energy dispersion relations. It was therefore suggested that
those measurements would be useful to reveal the Fermi surface
topologies.

Due to the complex nature of thermopower signals in anisotropic
materials, it is useful to briefly consider previous studies in
organic materials, including those at high magnetic fields where
field induced phase transitions occur. For consistency in the
anisotropic thermopower parameters reported below,
we use the following notation: the measured values can be written as
\tep[ij]{\alpha}, where $i, j = x$, $y$ or $z$,
signify  that the heat current is applied along a geometrical
direction $j$ and the thermal emf is measured along a direction $i$,
and $\alpha$ is one of the crystallographic axes along which the
heat current is applied. For zero magnetic field, \tep[ij]{\alpha}
can be written in the reduced form such as \tep[xx]{\alpha}, or
simply \tep[]{\alpha}.

Mori et al. \cite{mori2} performed the first in-plane measurements
on \Khg, where the temperature dependence of $S^a$ and $S^c$ were
found to be consistent with the topology of the hole and the
electron bands (see Fig. 1). Here, in general, the TEP is positive
when the thermal gradient is parallel to the open orbit sheets
(where the closed hole orbits dominate the signal), and negative
when the thermal gradient is perpendicular to the open orbit
sheets (where the electron orbits are the main contribution). Our
results, shown in Fig. 2 (and to be discussed below), are in
reasonable agreement with these measurements. This assignment is
supported by a simulation of the thermopower, based on Boltzmann
transport theory, with the bandwidth parameters as
input\cite{mori2,mori3}.

Another experiment involving in-plane thermopower has been
performed on the 10.4 K organic superconductor\cite{yu}  $\kappa$
- (BEDT-TTF)$_2$Cu[N(CN)$_2$]Br (see also Gartner et al.
\cite{gartner} and Buravov et al. \cite{buravov}). This system
also has open and closed orbit Fermi surface bands, and again, the
sign of the TEP in the normal state above $T_c$ follows the
general rule that it is positive when the thermal gradient is
parallel to the electron sheets, and negative when perpendicular.
Below $T_c$, the TEP vanished, as expected for a superconductor in
zero magnetic field.

In studies of quasi-one dimensional (Q1D) systems based on the
TCNQ charge transfer salts \cite{chaikin1, chaikin2} and the
perylene based compounds\cite{bender}, various features of TEP
were exhibited. Depending on the species of cations and the
temperature range, they show linear temperature dependence, 1/$T$
behavior, or temperature independence. The linear decrease of TEP
was treated by a simple, single carrier in 1D tight binding band
model, where the temperature behavior followed a linear $T$
dependence, $S = 2\pi^2k_B^2T/3eW~
\mathrm{cos}(\pi/2\rho)/\mathrm{sin}^2(\pi/2 \rho)$. Here $W$ is
the bandwidth and $\rho$ is the number of carriers per site. Below
the metal-to-insulator transition, the low temperature behavior
followed the 1/$T$ form $S = - (k_B/e) [(c-1)/(c+1)~E_g/2 k_BT +
3/4~ \mathrm {ln} (m_e/m_h)]$ where $E_g$ is the gap and $c$ is
the ratio of the electron to hole mobilities. The temperature
independent thermopower with magnitude of $\sim \pm60 \mu$V/K was
attributed to the spin configuration entropy ($\pm k_B/e
~(\mathrm{ln} 2)$) in the strong Coulomb potential limit ($U \gg
t$) in a quarter filled band.

In the presence of an applied magnetic field, the transverse
thermopower (the Nernst effect) may be measured in addition to the
normal longitudinal TEP (the Seebeck effect). Logvenov et al.
\cite{logvenov} have investigated the superconducting states of
two $\kappa$ - (BEDT-TTF)$_2$X materials up to 3 T. The results,
which showed non-zero TEP values even without magnetic fields,
were interpreted in terms of flux motion and the Magnus force (in
the case of the Nernst effect). The authors found considerable
anisotropy which was attributed to the difference in the electron
and hole Fermi surface sections.

The quasi-one dimensional Bechgaard salts, which exhibit spin
density wave (SDW) formation have been investigated by TEP and
MTEP in some detail. The SDW transition in (TMTSF)$_2$AsF$_6$ at
ambient pressure (T$_{SDW}$= 12 K) has been investigated
\cite{bondarenko} up to 11.3 T. Hysteretic temperature effects in
the thermopower were observed as a function of field direction
below T$_{SDW}$, and the authors speculated about pinning,
structural changes, and sub-phases associated with the SDW ground
state to describe the data. For systems with magnetic field
induced spin density wave states (FISDW), two thermopower studies
to 30 T have done on the quasi-one dimensional organic conductor
series (TMTSF)$_2$ClO$_4$ \cite{yu2} and (TMTSF)$_2$PF$_6$ under
pressure\cite{kang}. In these materials a FISDW transition occurs
from a Q1D metallic state to a sequence of SDW states. Both the
MTEP and Nernst signals, which bear some relation to the
corresponding transport measurements (i.e. longitudinal
resistivity $\rho_{xx}$ and the Hall effect $\rho_{xy}$
respectively), are sensitive to these transitions. As the final
FISDW sub-phase is entered at constant field with lower
temperatures, the thermopower first rises with an activated
(gap-like) behavior, followed by a decrease at lower temperatures
deeper in the SDW phase. The behavior is interpreted in terms of
collective modes within the SDW phases. Notably, quantum
oscillations (the so-called "rapid oscillations" which are
periodic in inverse field) are observed in the thermopower for
both materials. These experiments show the usefulness of
thermopower measurements to probe ground states that are induced
at high magnetic fields.

In this paper, we report thermopower measurement results carried
out under high magnetic field and at low temperatures in \Khg~.
The heat current was applied along all three crystallographic
axes, and the longitudinal (Seebeck coefficient) and the
transverse (Nernst-Ettinghausen effect) thermopower was measured.
The asymptotic behaviors of the Seebeck coefficient and the
Nernst-Ettinghausen effect suggest that the magnetic breakdown of
the closed orbit in the reconstructed Fermi surface are
responsible for the observed behavior of magnetothermopower.

\section{Experiment}
\Khg ~single crystal samples were obtained from the conventional
electrochemical crystallization technique. The orientation of
crystallographic axes were determined from polarized IR reflection
measurements at room temperature. Three different samples from a
single batch were used for each three different experimental
conditions (heat current $\parallel$ $a$-axis, $c$-axis or
$b$-axis). The magnetic field was always applied along the least
conducting axis ($b$-axis). When the heat current was applied in
the most conducting plane ($ac$-plane), off-diagonal term of
thermopower tensor (\tep [yx]{}) was also measured simultaneously.
The polarity of magnetic field was switched for the measurement of
\tep[yx]{} to remove spurious components due to the misalignment
of the voltage leads. (The alignment of voltage leads for
\tep[xx]{} measurements was also checked in this manner.)
Thermopower measurements were carried out by applying sinusoidal
heat currents along a crystallographic axis of the single crystal
and measuring thermal emf along the same direction and along the
direction rotated by 90 degrees. (see Fig. \ref{fig1} (c)) The
sample was mounted between two quartz blocks, which were heated by
sinusoidal heating currents (with oscillation frequency $f_0$ = 66
mHz) with $\pi$/2 phase difference. The corresponding temperature
gradient (\DelT) and thermal emf with 2$f_0$ oscillation frequency
were measured. The magnetothermopower of Au lead wires was in situ
calibrated by YBa$_2$Cu$_3$O$_{7+\delta}$ superconductor. To avoid
an ambiguity in determination of \DelT~ under magnetic field, we
exploit the reproducibility of \DelT, which only weakly depends on
magnetic field. The procedures of the magnetothermopower
measurement method used in this work are detailed
elsewhere\cite{eschoi}.

\section{Results and Discussion}
\subsection{Zero field thermopower}
The temperature dependence of zero field thermopower results For
\Khg, along each of the principal axes is shown in
Fig.~\ref{fig2}, where the inset shows low temperature expansion
near $T_p$ = 8 K. The in-plane results are in general agreement
with previous measurements by Mori et al.\cite{mori2}, as
discussed in Sect. II. The negative (positive) sign of thermopower
along the $a(c)$-axis indicates that the Fermi surface is electron
(hole)-like along the $a(c)$-axis, which corresponds to the Fermi
surface as shown in Fig. 1: here, for a thermal gradient along the
$c$-axis direction, the hole pockets should contribute, whereas
for the $a$-axis direction, the electron-like open orbits should
contribute as well. We note however, that the temperature
dependence of the TEP shown in Fig.~\ref{fig2} is non-monotonic.
There is a minimum in \tep[]{c} and \tep[]{a} around T = 120 to
150 K .This behavior can be explained by applying the
two-dimensional tight binding parameters for this material to the
Boltzmann transport equation\cite{mori2},\cite{mori3}.

The inter-plane TEP (\tep[]{b}), perpendicular to the conducting
planes, which has not been previously measured, shows a monotonic
decrease until 35 K, where it changes sign. The positive sign of
\tep[]{b} above $T$ = 35 K indicates that the hole-like carriers are
dominant for the inter-layer transport above $T$ = 35 K and the
electron-like carriers dominate below that temperature. Clearly,
the inter-plane transport shows a distinctly different behavior to
that of the in-plane TEP.

 From the linear dependence of \tep[]{c} and \tep[]{b} at higher
temperature, the density of states per carriers are derived to be
5.3 states/eV for \tep[]{c} and 7.6 states/eV for \tep[]{b}. The
deviation from the linear dependence in the intermediate
temperature range has been treated\cite{mori2,mori3} by Mori et
al., as mentioned in Sect. II above, as has the peak in \tep[]{c}
near 25 K . The peak in \tep[]{c} may arise from the
energy-dependent scattering term of Eq. \ref{mott}, or from  the
phonon drag effect. The TEP associated with phonon drag results
from the electron-phonon coupling. The temperature dependence
changes from a $T^3$ to a $T^{-1}$ behavior \cite{chaikin} with
increasing temperature. Especially, peak structures at low
temperature (typically between $\theta_D$/10 and $\theta_D$/5,
where $\theta_D$ is the Debye temperature) are usually attributed
to the phonon drag effect, whose temperature dependence in noble
metals is quite similar to that of \tep[]{c}. Thermopower
measurements on the other members of this compound
($\alpha$-(BEDT-TTF)$_2$MHg(SCN)$_4$, where M=Tl, Rb, NH$_4$) also
showed peak structures at lower temperatures (15 $\sim 20$ K),
which were attributed to the phonon drag effect\cite{demish}. Of
note is the superconductivity of the NH$_4$ salt, where there is
an enhancement of density of states and effective mass (m*
$\approx 2.1 m_0$ for $\kappa$-(BEDT-TTF)$_2$NH$_4$Hg(SCN)$_4$
over that of non-superconducting \Khg~ (m* $\approx 1.5 m_0$)), as
pointed out by Mori et al.\cite{mori2} However, the origin of
enhancement of m*, which may be due to electron-electron
interaction and/or electron-phonon interaction, is not yet clear.

  When the heat current is applied perpendicular to the most conducting planes,
the 2D band model will not be appropriate without considering the
energy dispersion between conducting planes. Since simple mixing
of two contributions of \tep[]{a} and \tep[]{c} cannot produce
temperature dependence and magnitude of \tep[]{b}, it is necessary
to find a proper model for inter-plane transport. A
straight-forward application of the $S \approx T/W $ relationship
produces unrealistic values of the $b$-axis bandwidth (of order
250 meV), which is known from Fermiology studies to be at least
100 times smaller. At present there is no satisfactory treatment
of either the magnetoresistance or the thermopower for the
$b$-axis transport behavior.

The behavior of the thermopower through the density wave
transition ($T_p$= 8 K) has also not been previously reported.
From the inset of Fig. \ref{fig2}, jumps of thermopower for
\tep[]{a} and \tep{c} are observed near the metal-density wave
transition temperature ($T_p$). In the resistance measurements,
the change of slope near $T_p$ may be attributed to the vanishing
conductivity of the open orbit band ($\sigma_{1D}$) in parallel
with the remaining finite conductivity of closed orbit band
($\sigma_{2D}$)\cite{brooks}. In addition, for thermopower
measurements, metallic carriers give a linear $T$-dependence while
activation over a band gap gives a 1/$T$ dependence. Therefore for
one type of carrier (either from open orbit or closed orbit)
exclusively, \tep[]{a} will diverge below $T_p$. Experimentally,
there is always the possibility of random diffusion of heat along
the sample due to misalignment between heat currents and crystal
axes, and also, the total thermopower may involve a mixing of
contributions from several bands (especially below $T_p$ where the
Fermi surface is reconstructed). Nevertheless, we believe that
dominant behavior of thermopower in the present case reflects the
influence of the heat currents with respect to the topology and
orientation of the specific orbits.

When two types of carriers are involved, the total thermopower can be
  written as\cite{mcdonald},
\begin{equation}
\label{Stot} S_{tot} =
\frac{\sigma_{1D}S_{1D}+\sigma_{2D}S_{2D}}{\sigma_{1D}+\sigma_{2D}}
\end{equation}
Above $T_p$, both orbits have metallic conductivity and
thermopower; hence the total thermopower shows a linear
temperature dependence. But below $T_p$, a band gap opens for $1D$
band, therefore $\sigma_{1D}$ goes to zero exponentially as the
temperature decreases, and $S_{1D}$ diverges as 1/$T$. However,
below this temperature, a modified closed orbit band remains,
which still gives a metallic thermopower contribution.
Consequently, just below the transition temperature, there will be
a jump of thermopower, $\Delta S$, from the divergence of $S_{1D}$, but this
contribution quickly disappears as $\sigma_{1D}$ goes to zero.

We may further compare the magnitude of jump of
thermopower ($\Delta S$) and that of electronic heat capacity
($\Delta C_{el}$). Since thermopower measures heat carried by a
carrier divided by the carrier charge and temperature, $\Delta S$
equals $\Delta C_{el}$/$e$. Adding both contributions from
\tep[]{a} and \tep[]{c}, $\Delta S$ is about 1 $\mu$V/K, which
corresponds to $\frac{1}{6.24}$ $\times 10^{-24}$ J/
K$\cdot$carrier. The reported $\Delta C_{el}$ is about 0.1
J/mol$\cdot$K\cite{kovalev} and 0.25 J/mol$\cdot$K\cite{henning},
which are in fair agreement with $\Delta S$. A corresponding change
in $S^b$ at $T_p$ is unclear, if it exists
at all (in Fig. 2). This indicates that the phase transition around
$T_p$ = 8 K occurs
predominantly within the in-plane electronic structure.

\subsection{Magnetothermopower}
In Fig. \ref{fig3}, we present the magnetic field dependence of
the TEP (i.e. MTEP) for all three crystallographic directions at
different, fixed temperatures. To show the field dependence more
clearly, some traces are offset from zero as indicated by the
dashed lines. The MTEP signal contains two components, the
background MTEP which is sensitive to the Fermi surface topology,
and the oscillatory MTEP which is a manifestation of the Landau
quantization of the closed orbit Fermi surface. To separate the
two signals, we show in Fig. \ref{fig4} the background MTEP data
where the oscillatory MTEP has been removed by filtering the total
MTEP signal. Finally, the Nernst-Ettinghausen effect at different
temperatures is shown in Fig. \ref{fig5}. In all cases the
magnetic field is applied along the $b$-axis, perpendicular to the
$a-c$ conducting plane.

A similarity between the magnetic field
dependence of the thermopower and the resistance can be expected
from the Boltzmann transport equations. Indeed we find that the
general field-dependence of the MTEP is similar in some respects
to that of the magnetoresistance\cite{sasaki1}: (1) Below $T_p$,
the in-plane MTEP shows a broad peak at $B (= B_A) \sim$ 10 T. (2)
Above the "kink field" $B_k$ at 22.5 T, the field dependence is
weak. (3) Quantum oscillations are observed in the MTEP signal
associated with the de Haas-van Alphen (dHvA) and Shubnikov - de
Haas (SdH) effects (i.e., from Eqn. 1, both thermodynamic and
transport properties are involved in the thermopower). The
oscillation amplitudes are largest for \tep[xx]{c}, and the
oscillations are also observed in \tep[zz]{b}. (4) The MTEP also
shows hysteresis for up and down field sweeps.

Beyond the general comparison between resistivity and MTEP, unique
field-dependent features are observed in the MTEP and Nernst signals:
(1) For
\tep[xx]{a} and \tep[xx]{c}, the MTEP exhibits a minimum at
$B=B_{min}$ after which it rapidly increases up to maximum at
$B=B_A$. $B_{min}$ is also temperature dependent, and decreases with
increasing temperature.

(2)For the $a$-axis behavior, the Nernst signal \tep[yx]{a} shows
a similar field dependence to the MTEP signal \tep[xx]{a} in
general, but \tep[yx]{a} is asymptotic to zero above $B_k$ for all
temperatures measured. In contrast, \tep[xx]{a} is asymptotic to
non-zero, temperature dependent values above $B_k$.

(3) For the $c$-axis behavior, \tep[yx]{c} shows linear field
dependence at higher field, which is distinct from the other
thermopower coefficients.

(4) For the $b$-axis, \tep[zz]{b} exhibits some aspects of
\tep[xx]{a} and \tep[xx]{c} mentioned above, which may involve
some mixing of the in-plane components.

Fig. \ref{fig6} shows the temperature dependence of
magnetothermopower for fixed magnetic field. As the magnetic field
is increased, the transition temperature $T_p$, represented as a
jump or shoulder in the data,  shifts to lower temperature. (For
\tep[zz]{b} and \tep[xx]{c} at $B$=27 T however, $T_p$ could not
be resolved.) The temperature-field phase diagram, based
exclusively on the  field and  temperature dependent thermopower
data ($B_{min}, B_A, B_K$, and $T_p(B)$), is shown in Fig.
\ref{fig7}. The phase diagram is very similar (with the exception
of the new $B_{min}$ line) to previous determinations by
transport\cite{sasaki,biskup}, NMR\cite{kanoda,kuhns},
  magnetization\cite{harrison1,christ}, and
specific heat\cite{kartsovnik3}.

\subsubsection{Theoretical aspects of high field thermopower measurements}
To interpret the origin of the field dependence of the MTEP, it is
useful to examine the magnetic field dependence of the thermopower
for an anisotropic system. When thermopower is measured under
magnetic field, tensor elements of thermopower should be
considered since both the temperature gradient and the magnetic
field are vectorial. For our experimental conditions (zero
electric current and zero transverse heat current), the kinetic
equations for thermopower and heat currents can be written as,

\begin{alignat}{3}
\label{kinetic}
  E_i = \alpha_{ij}\nabla T_j,
  & \qquad (J_q)_i = -K_{ij}\nabla T_j
\end{alignat}

where $E$ is the electric field produced by the thermal emf,
$\alpha$ is the thermopower tensor under magnetic field, $J_q$ is
the heat current, and $K$ is the thermal conductivity tensor.
We have two different experimental situations.

(a) $H$ $\parallel$ $z-$ axis $\parallel$ $J_q$ $\parallel$
$\nabla T_z$ (\tep[zz]{b} in this paper) : In this case, no
Lorentz force is applied so that \DelT$_x$= \DelT$_y$ = 0, and
above equations have simple scalar form,
\begin{alignat}{3}
\label{Szz}
  E_z = \alpha_{zz}\nabla T_z,  & \qquad
(J_q)_z = -K_{zz}\nabla T_z
\end{alignat}
The measured value \tep[zz]{b} is obtained from $E_z$/\DelT$_z$.

(b) $H$ $\parallel$ $z-$ axis $\perp$ $J_q$ (\tep[xx]{a},
\tep[yx]{a}, \tep[xx]{c} and \tep[yx]{c} in this paper) : In this
case, \DelT ~is not parallel with $J_q$, and has a non-zero $y$
component given by  -\DelT$_x$$K_{yx}$/$K_{yy}$. If we rewrite Eq.
\ref{kinetic} for this case,

\begin{equation}\label{Sxy}
\begin{pmatrix}
   E_x \\
   E_y
\end{pmatrix}
=
\begin{pmatrix}
   \alpha_{xx} & \alpha_{xy} \\
   \alpha_{yx} & \alpha_{yy}
\end{pmatrix}
\begin{pmatrix}
   \nabla T_x \\
   -\nabla T_x K_{yx}/K_{yy}
\end{pmatrix}
\end{equation}

In terms of measured values, Eq. \ref{Sxy} can be expressed as,
\begin{eqnarray}
\label{TEPmeasured} \nonumber \
  S_{xx}=\frac{E_x}{\nabla T_x}=\alpha_{xx}-\alpha_{xy}\frac{K_{yx}}{K_{yy}}
  \end{eqnarray}
\begin{eqnarray}
  S_{yx}=\frac{E_y}{\nabla T_x}=\alpha_{yx}-\alpha_{yy}\frac{K_{yx}}{K_{yy}}
\end{eqnarray}
The asymptotic behavior of thermoelectric tensor
$\alpha_{ij}$\cite{BGN} and thermal conductivity tensor
$K_{ij}$\cite{azbel} were calculated at low temperature and high
magnetic field. They show different behavior for the three different
cases, i.e., (a) closed and compensated orbits (b) closed and
uncompensated orbits and, (c) open orbits along the $x-$direction.
For each case, the asymptotic behaviors of \tep[ij]{} can be
calculated and the results are summarized in Table \ref{table1}.

The saturation values are temperature dependent for each case, and
they should be determined by considering the scattering and the
dispersion relations.

\subsubsection{Comparison of theoretical and experimental MTEP results}
    We now compare the thermopower data with the theoretical
field dependence described above. We focus our attention on the
behavior of the \tep[]{c} signal shown in Fig. 3 and 4, since it
is significantly larger than the other signals, and it is
predominantly hole-like (i.e. positive, and closed
orbit-sensitive). We may use as a guide the detailed work by Uji
et al. on the Fermiology study of the quantum
oscillations\cite{uji}, due to closed orbits in the magnetic
breakdown network, which have been proposed to describe the Fermi
surface topology below $T_p$. In this model it is assumed that
only small closed orbit pockets exist in the reconstructed Fermi
surface below $T_p$ near zero magnetic field. As the magnetic
field is increased, larger and larger Fermi surface sections
become involved in the SdH oscillations as magnetic breakdown
becomes more probable. These effects appear experimentally in the
SdH measurements: in Figs. 8 and 9 of Ref. [45], the onset of the
SdH oscillations associated with the various magnetic breakdown
orbits ($\alpha$, $\phi$, etc.) occurs above 3.5 T. At higher
fields, above 7 T, the $\beta$ orbit associated with the Brillouin
zone appears, i.e. magnetic breakdown occurs over most of the
Fermi surface. Finally, above the kink field, the original Fermi
surface topology (as in Fig. 1) re-emerges.

    In light of the above, our first point of comparison is the
behavior of $B_{min}$. For \tep[]{c}, and indeed in all MTEP
directions, there is a distinct change in the signal above
$B_{min}$ from a relatively flat response to a pronounced increase
in slope (approximately linear) with field. This behavior is
entirely consistent with the theoretical model above for a set of
closed, compensated orbits.

At higher fields, above $B_A$, the compensated
closed orbit behavior of \tep[]{c} changes dramatically, and this
indicates that the magnetic breakdown network of closed orbits is
undergoing modification with increasing field.  By comparison with
the theory above, we may speculate that this is the onset of
uncompensated closed orbit behavior, which exhibits less
(eventually no) field dependence.

Finally, above the kink field, the original, uncompensated closed
orbit FS is realized, and the field dependence disappears, as is
evident from the \tep[]{c} data. Above $B_K$, we may compare the
temperature dependence of \tep[]{c} with those at zero field above
$T_p$ (See Fig. \ref{fig6}). Except for field dependent offsets,
the thermopower above $T_p(B)$ for the three directions is
similar, save for the case of \tep[]{b} which is anomalous, even
in the absence of field.

Beyond this simple comparison of the most dominant MTEP signal
with previous SdH studies and theoretical expectations, the
assignment the mechanisms responsible for the magnetic field
dependence of \tep[]{a} and \tep[]{b} are more speculative. This
is due to the apparent mixing of terms in the experimental data,
and to sort them out is beyond the scope of the present work.
However, the  Nernst data shows significant difference for the $a$
and $c$ axes (\tep[yx]{a} and \tep[yx]{c}). This is most apparent
above $B_k$, where the original FS is expected to be recovered.
Here \tep[yx]{a} is seen to be asymptotic to zero (i.e. $H^{-1}$
dependence), whilst \tep[yx]{c} clearly assumes a linear (i.e.
H$^1$) dependence. This behavior is consistent with expectations
based on the theory represented in Table 1.

\subsubsection{MTEP and Landau quantization}

 From Eq. 1, it is clear that for a quasi-two dimensional Fermi
surface, the magnetic field dependence of the Landau level
spectrum\cite{LK} will produce quantum oscillations in the
thermopower, which will originate from both thermodynamic and
electron transport factors. The latter has been treated by
Zil'berman\cite{zilberman} and Long et al. \cite{long} where
scattering between states of the Landau level closest to the Fermi
surface and states of other occupied are considered. We find in our
results a clear indication of the Landau level spectrum in the MTEP
and Nernst data, the periodicity of which is in good agreement with
previous SdH and dHvA measurements\cite{wosnitza}. In particular,
the oscillation frequency obtained from the fast Fourier transform
(FFT) analysis
 of \tep[xx]{c} and \tep[yx]{c} at 0.7 K (see Fig. \ref{fig8})
is about 667 T, which agrees with results obtained from
other measurements\cite{uji}. This comparison also includes the
observation of higher harmonics of the fundamental $\alpha$ orbit,
and even the $\phi$ orbit (180 T) which is expected to be a
manifestation of the magnetic breakdown topology below $B_k$. (See
for instance Uji et al.\cite{uji}).

An important feature of our results, most evident above $B_k$, is
a phase difference between the METP and Nernst quantum
oscillations. This is shown in Fig. \ref{fig9} for two different
cases, the $a$-, and $c$-axes.  Indeed,  Zil'berman's
model\cite{zilberman, long} predicts a phase difference of $\pi$
between the MTEP and Nernst oscillations. We show this more
explicitly for \tep[xx]{c} and\tep[yx]{c} at $T$=0.7 K in Fig.
\ref{fig10} , with the angle ($\theta$) defined as the ratio of
the electric fields due to transverse and longitudinal thermopower
($\theta$=arctan ($S_{yx}/S_{xx}$)= arctan($E_y/E_x$). The phase
difference is $\pi$, which supports Zil'berman's
model\cite{zilberman}.

Zil'berman's model also predicts similar field and temperature
dependence of oscillation amplitude with that of the
Lifschitz-Kosevich formula for de Haas-van Alphen oscillations.
The amplitude of thermopower oscillation increases when $B_A < B <
B_k$, but it is almost field independent above $B_k$.  We note
that in this range of temperature, the quantum oscillation
amplitude is complicated due to the anomalous high field state,
where the dHvA and SdH behave in a significantly different
manner, and a comparison based on standard LK theory is not
applicable. A more systematic study of the MTEP and Nernst effects
above $B_k$ vs temperature, and also vs. field orientation, will
be necessary to provide a complete picture.

\section{Conclusion}

We have applied the method of magnetothermopower to investigate
the low temperature, magnetic field dependent phase of the
anisotropic organic conductor \Khg. Our results show that the
sensitivity of thermopower to anisotropic Fermi surface topologies
may be extended to the study of systems where temperature and
magnetic field alter these topologies. The most significant
results of the present investigation include the determination of
the low temperature phase diagram of \Khg , based purely on
thermopower measurements, the study of the onset of magnetic
breakdown effects via the magnetothermopower, and the measurement
of quantum oscillations in the magnetothermopower and Nernst
effect, which are in accord with theoretical expectations.

\begin{acknowledgments}
This work was carried out at the National High Magnetic Field
Laboratory, which is funded by a cooperative agreement between the
National Science Foundation and the State of Florida through
NSF-DMR 00-84173. JSB acknowledges support from NSF-DMR-99-71474
for this work, and ESC acknowledges post-doctoral support from
KOSEF.
\end{acknowledgments}

\newpage

\begin{table}
\caption{High field power law behavior ($H^\gamma$ where $\gamma$
= 1,0, -1) of the magnetothermopower for different types of orbits
\label{table1}}
\begin{tabular}{cccc}
&closed&closed&open along the\\
&compensated&uncompensated&$x-$direction\\ \colrule
\\$K_{yx}/K_{yy}$&$H^0$&$H^1$&$H^{-1}$\\
$\alpha_{xx},\alpha_{yy}$&$H^1$&$H^0$&$H^0$\\
$\alpha_{xy}$&$H^1$&$H^{-1}$&$H^1$\\
$\alpha_{yx}$&$H^1$&$H^{-1}$&$H^{-1}$\\
$\alpha_{zz}$&$H^0$&$H^0$&$H^{-1}$\\
\\ \tep[xx]{}&$H^1$&$H^0$&$H^0$\\ \tep[yx]{}&$H^1$&$H^1$&$H^{-1}$\\
\tep[zz]{}&$H^0$&$H^0$&$H^0$\\
\end{tabular}
\end{table}

\begin{figure}[tbp]
\caption{Fermi surface of \Khg~ (a) prior to and (b) after
reconstruction. {\bf q} denotes nesting vector. (c) Lead wire
configuration for magnetothermopower and Nernst effect
measurements.} \label{fig1}
\end{figure}
\begin{figure}[bp]
\caption{Temperature dependence of zero field thermopower of \Khg~
when the temperature gradient is applied along the $a$-axis
($S^a,*$), $c$-axis ($S^c,\circ$) and $b$-axis ($S^b,\triangle$).
The inset shows the low temperature behavior near the transition
temperature $T_p$.} \label{fig2}
\end{figure}
\begin{figure}[bp]
\caption{Magnetothermopower of (a) $S^a$, (b) $S^c$ and (c) $S^b$
for field down sweeps. Field sweep up data are also shown for T
$\sim$ 1.5 K. The arrows indicate characteristic fields as defined
in the text. Shifted zeros for some data sets are indicated by
dashed lines. For unshifted data, see Fig. \ref{fig4}}
\label{fig3}
\end{figure}
\begin{figure}[bp]
\caption{Background magnetothermopower of \tep[]{a} and \tep[]{b}
obtained from Fig. \ref{fig3} by filtering out the quantum
oscillation component.} \label{fig4}
\end{figure}
\begin{figure}[bp]
\caption{Nernst effect of \Khg when the heat current is applied
along $a$-axis (\tep[yx]{a}) and $c$-axis (\tep[yx]{c}).
}\label{fig5}
\end{figure}
\begin{figure}[bp]
\caption{Temperature dependence of magnetothermopower of \Khg~
measured under fixed magnetic field. Arrows indicate the
transition temperature $T_p$.} \label{fig6}
\end{figure}
\begin{figure}[bp]
\caption{$T$-$B$ Phase diagrams of \Khg~ drawn from \tep[xx]{a}
and \tep[xx]{c}. $T_p$ is obtained from the temperature sweep and
$B_{min}$, $B_A$ and $B_K$ are obtained from the field sweep.
Empty symbols are from \tep[xx]{a} and filled symbols from
\tep[xx]{c}.} \label{fig7}
\end{figure}
\begin{figure}[bp]
\caption{FFT spectrum for \tep[xx]{c} and \tep[yx]{c} at $T$=0.7 K
and 1.5 K for 8.6 T $\leq B \leq$ 12 T and 8.6 T $\leq B \leq$ 27
T. The fundamental frequency of the closed hole orbit ($\alpha$
orbit) and its higher harmonics are clearly shown. The peak at 180
T is the $\phi$ orbit, as in Ref. \cite{uji} } \label{fig8}
\end{figure}
\begin{figure}[bp]
\caption{Transverse and longitudinal magnetothermopower
oscillation as a function of inverse magnetic field in the high
field region for (a) \tep[]{a} and (b) \tep[]{c}. The background
magnetothermopower was subtracted and the amplitude of \tep[xx]{c}
was divided by 5.} \label{fig9}
\end{figure}
\begin{figure}[bp]
\caption{Comparison of the phases of oscillations between
\tep[xx]{c} and \tep[yx]{c} at $T$= 0.7 K and plot of
arctan(\tep[xx]{c}/\tep[yx]{c}) as a function of magnetic field.}
\label{fig10}
\end{figure}
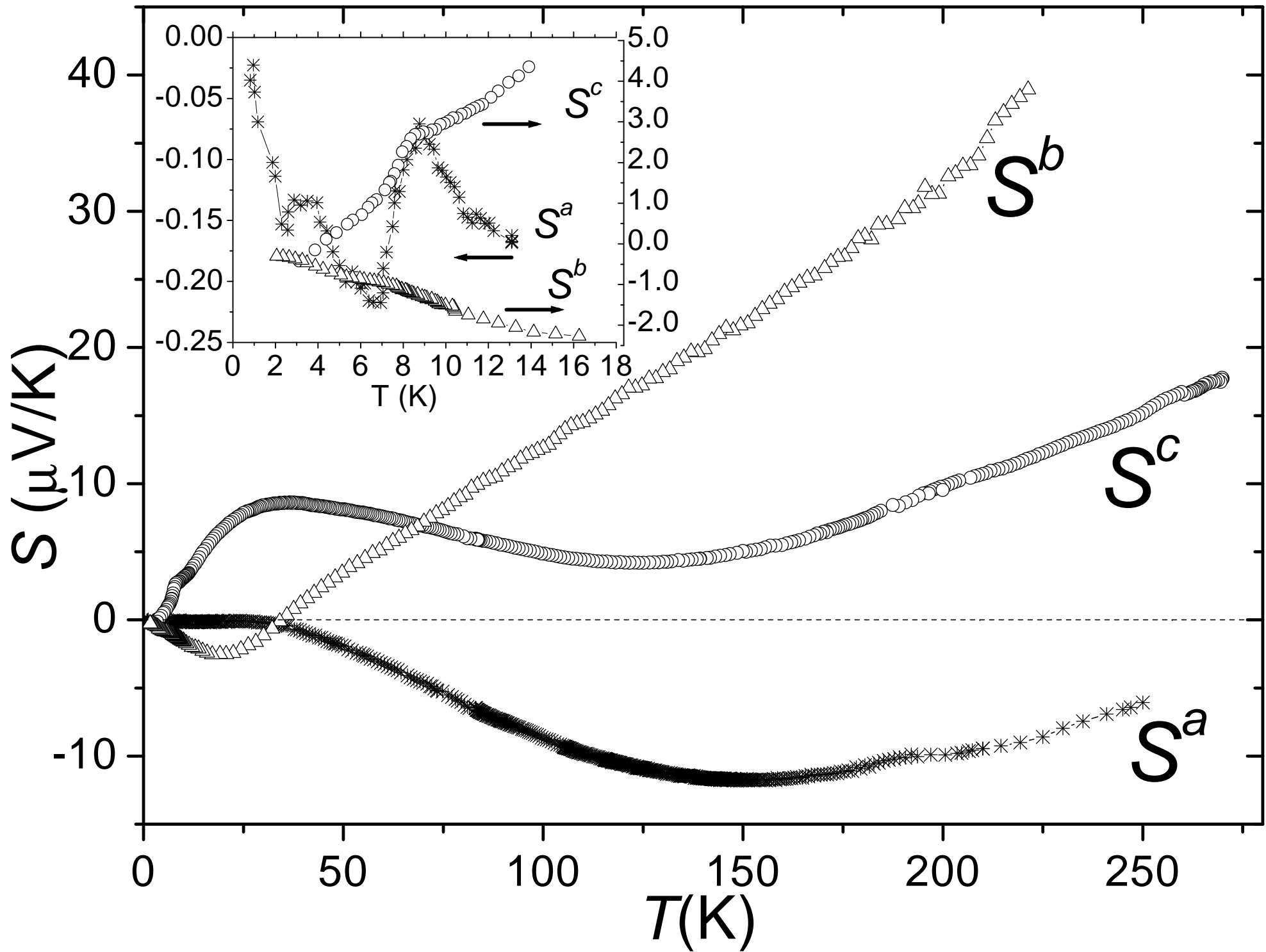
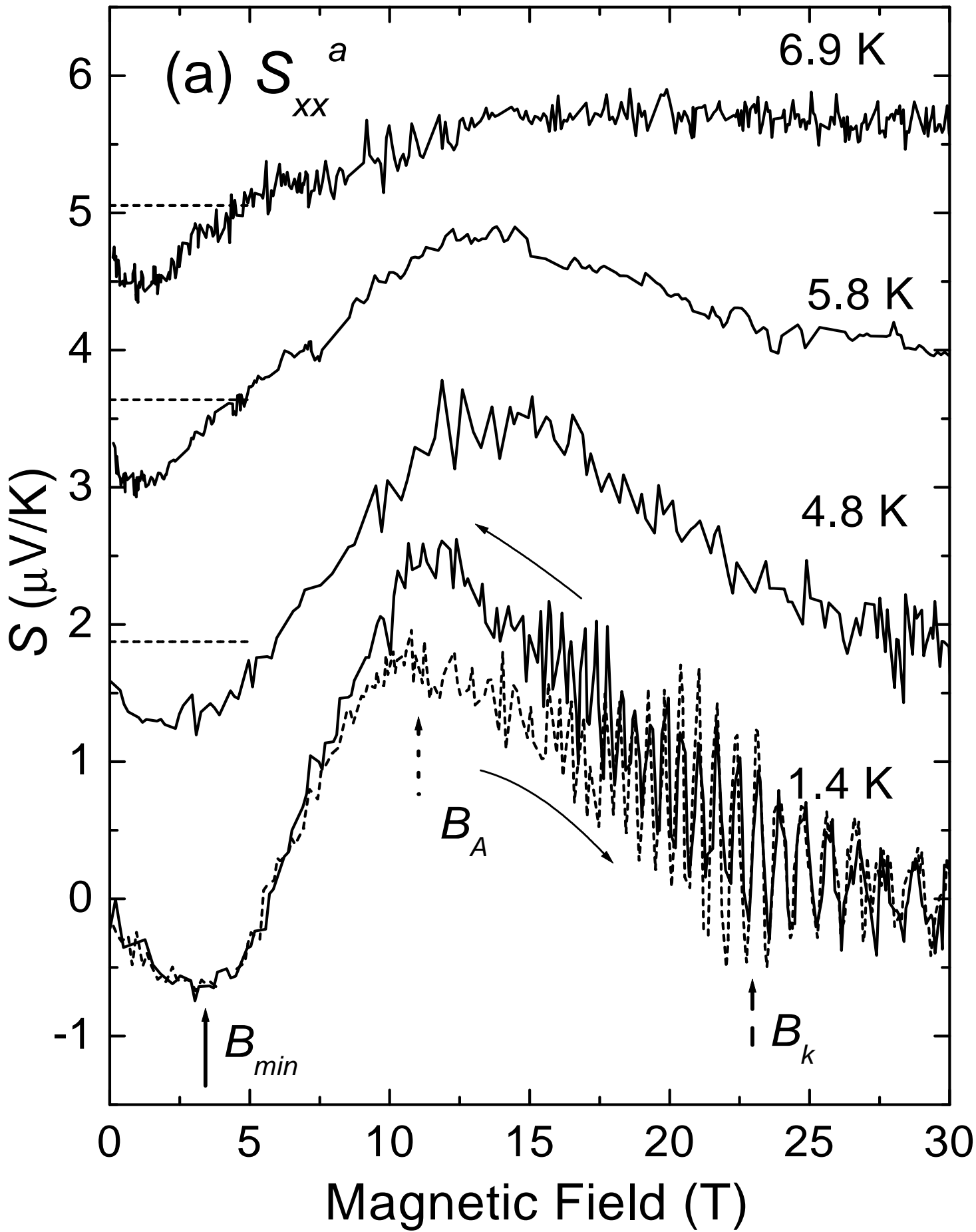
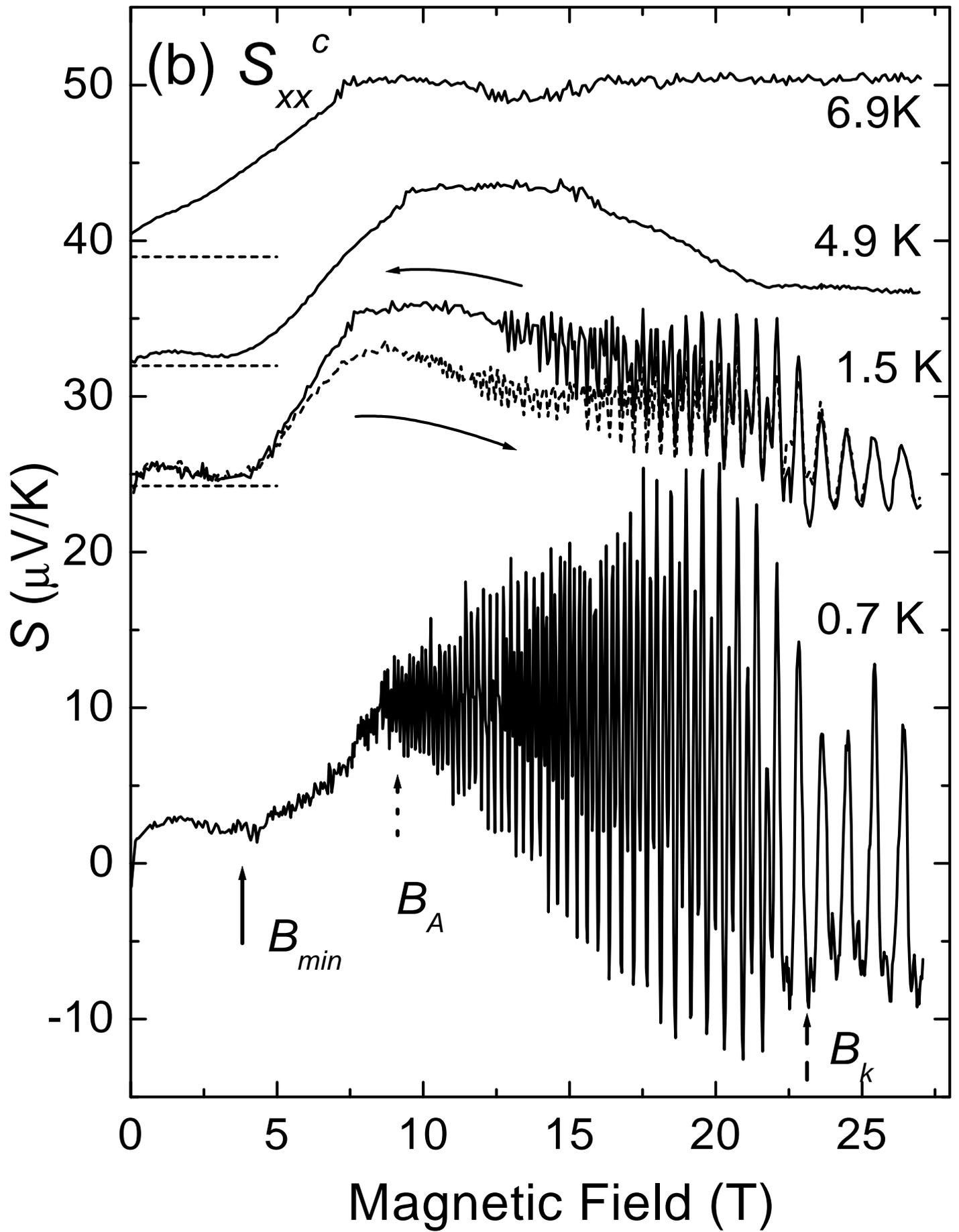
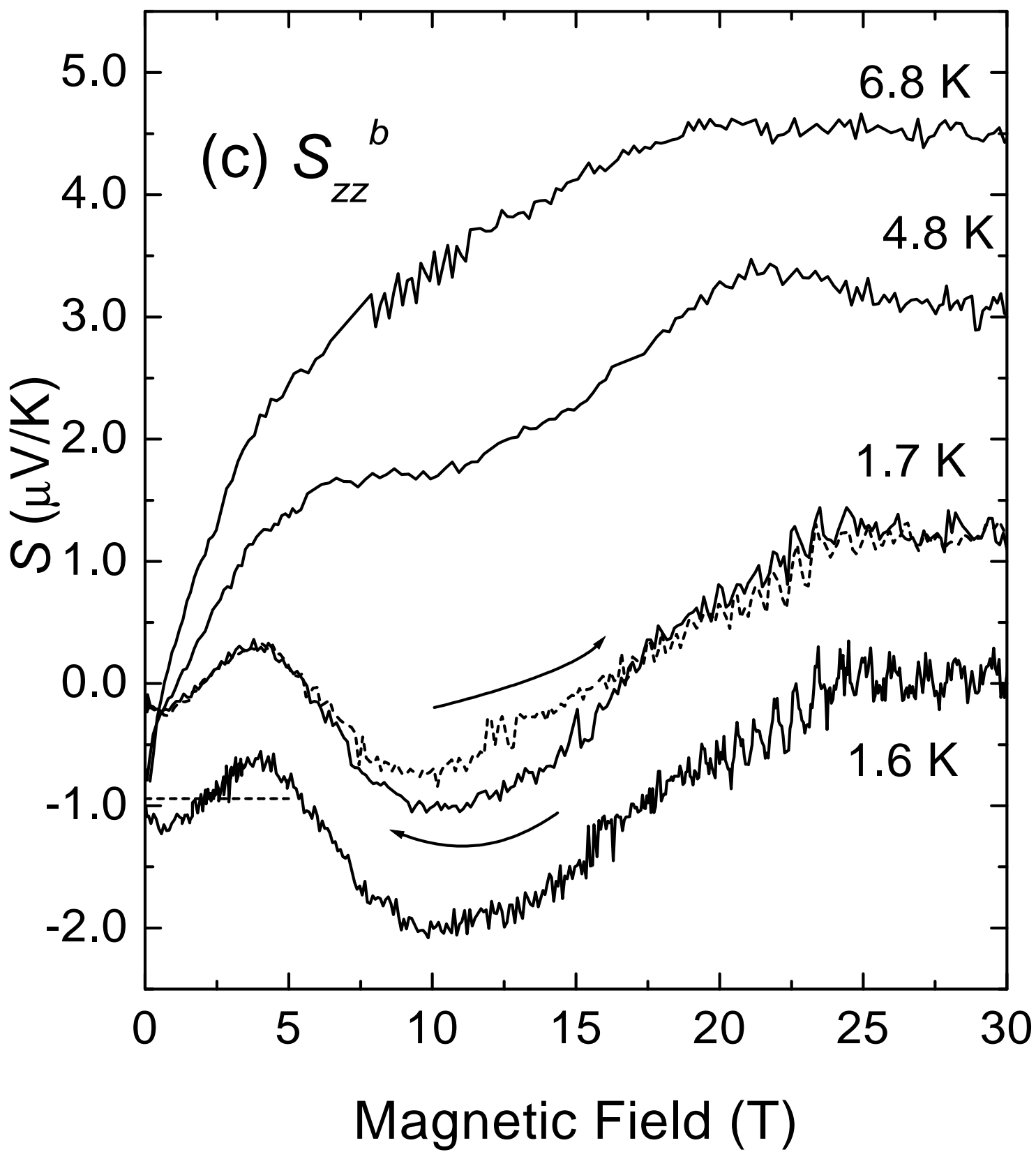
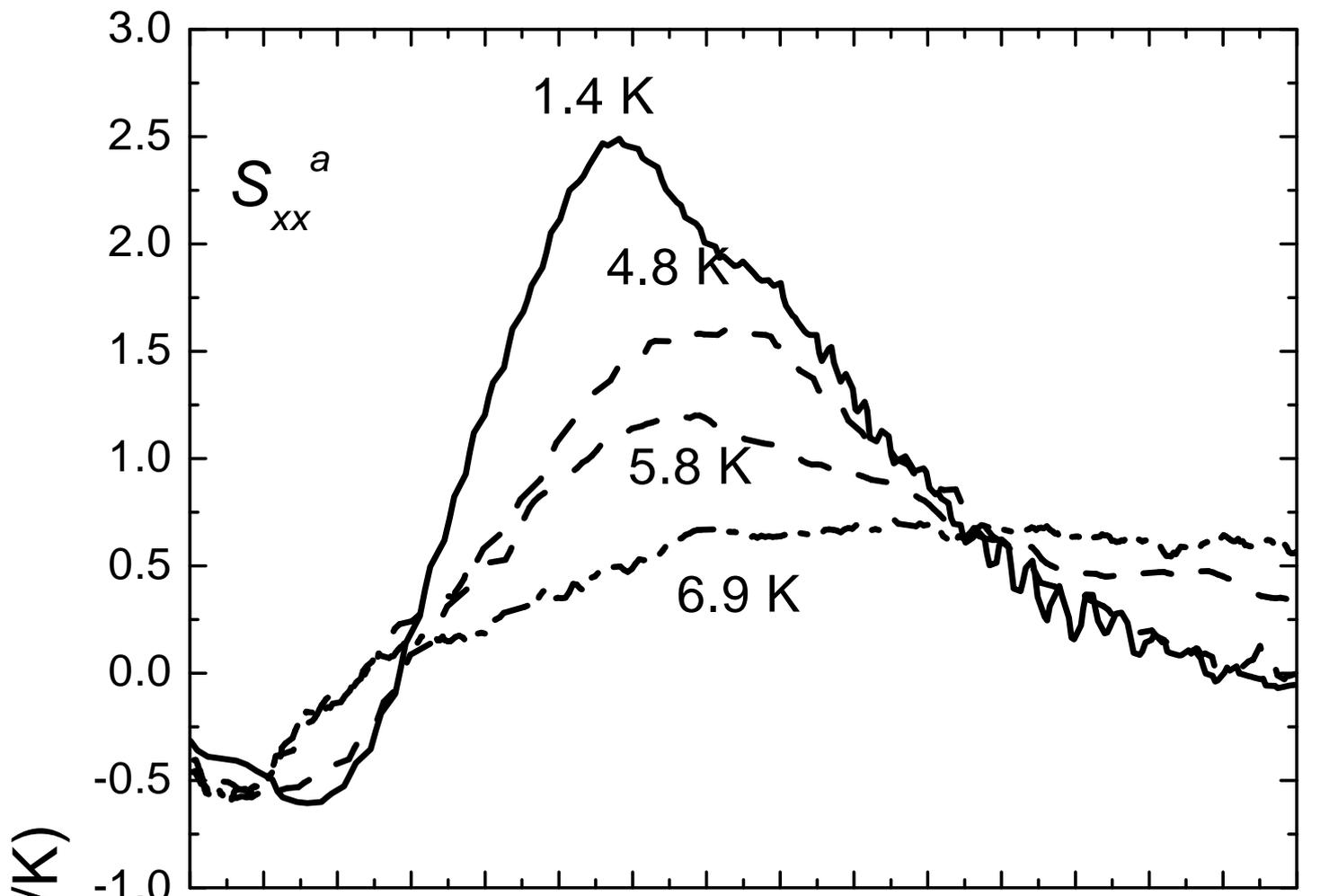
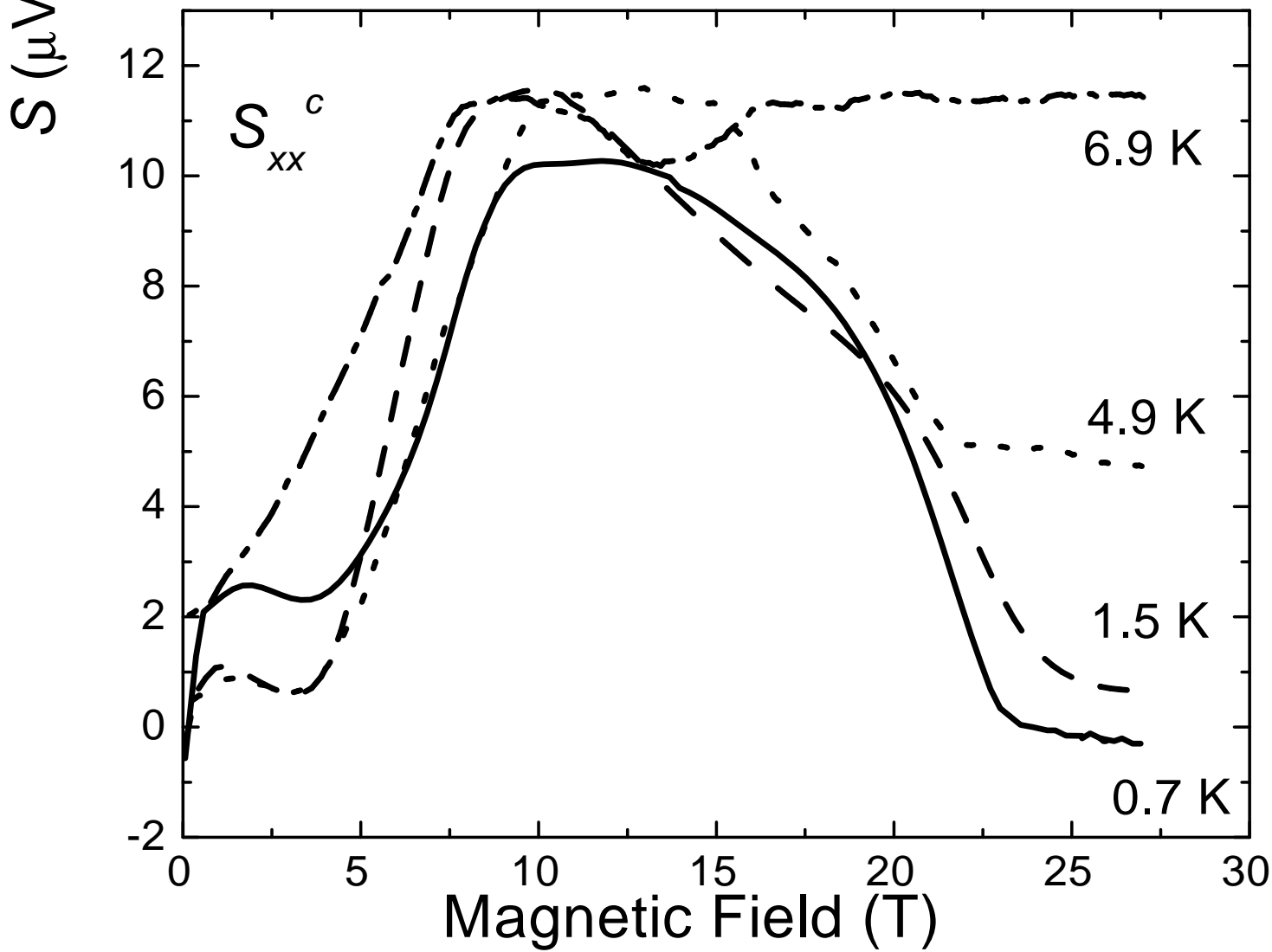
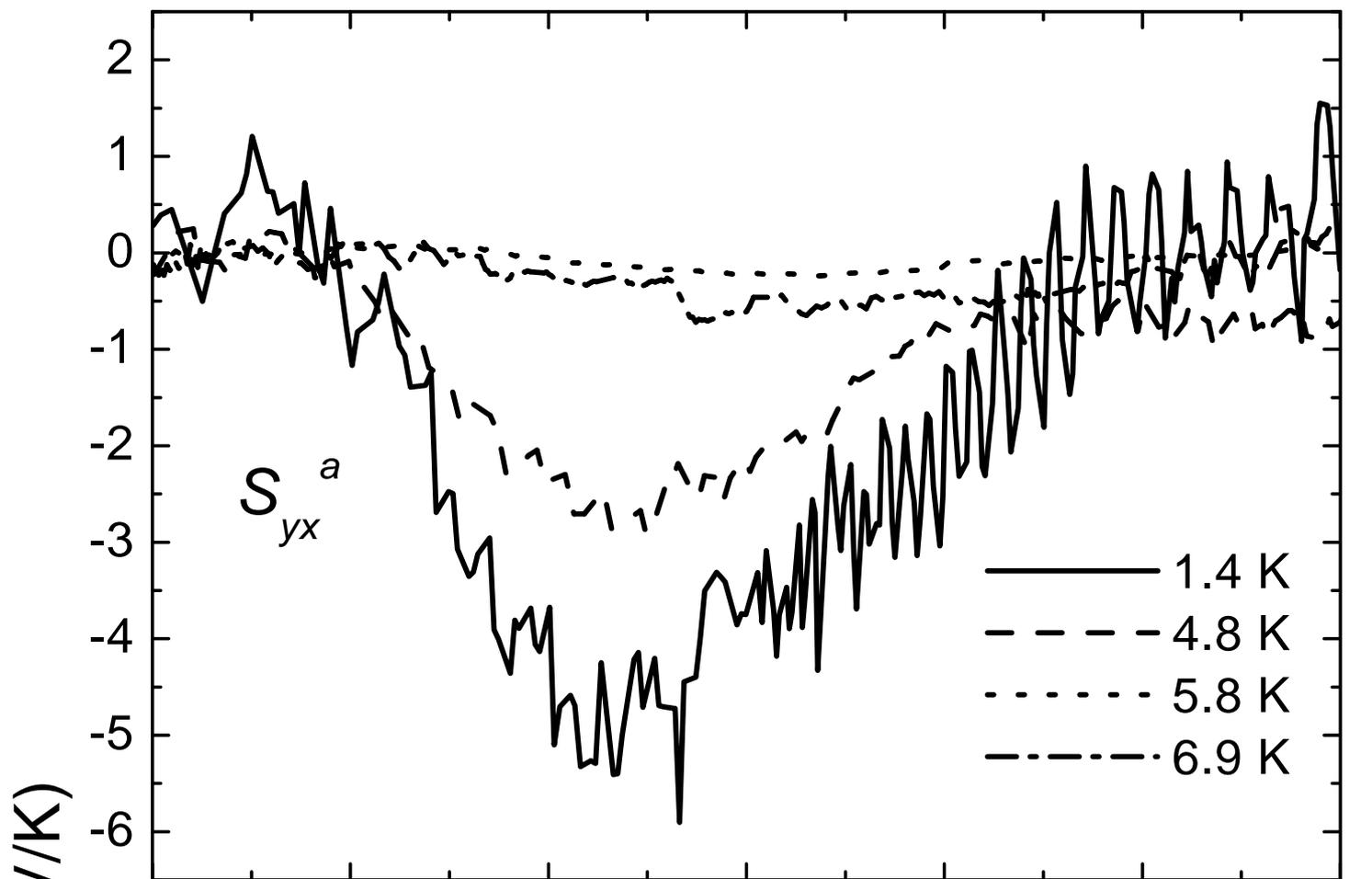
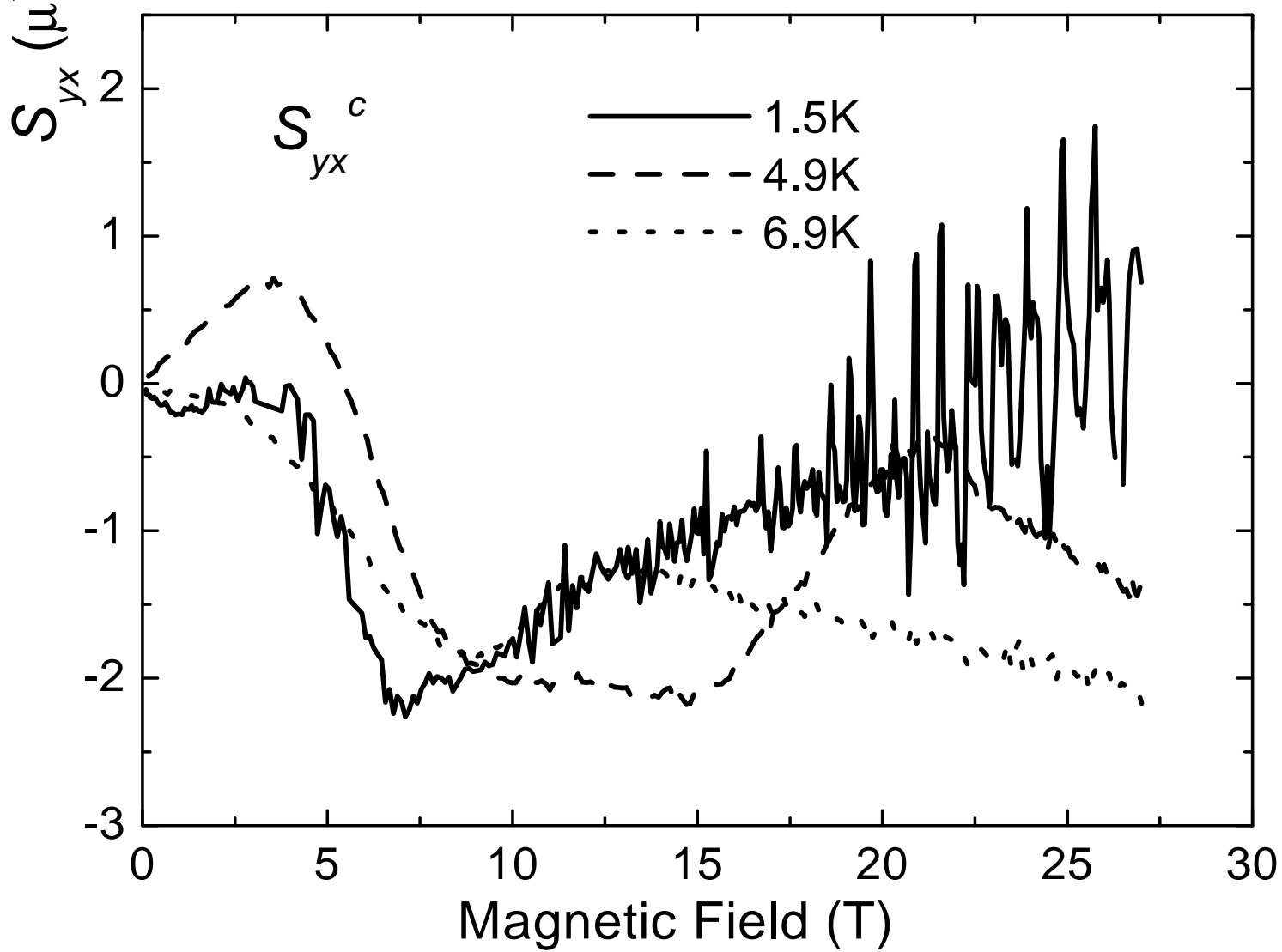
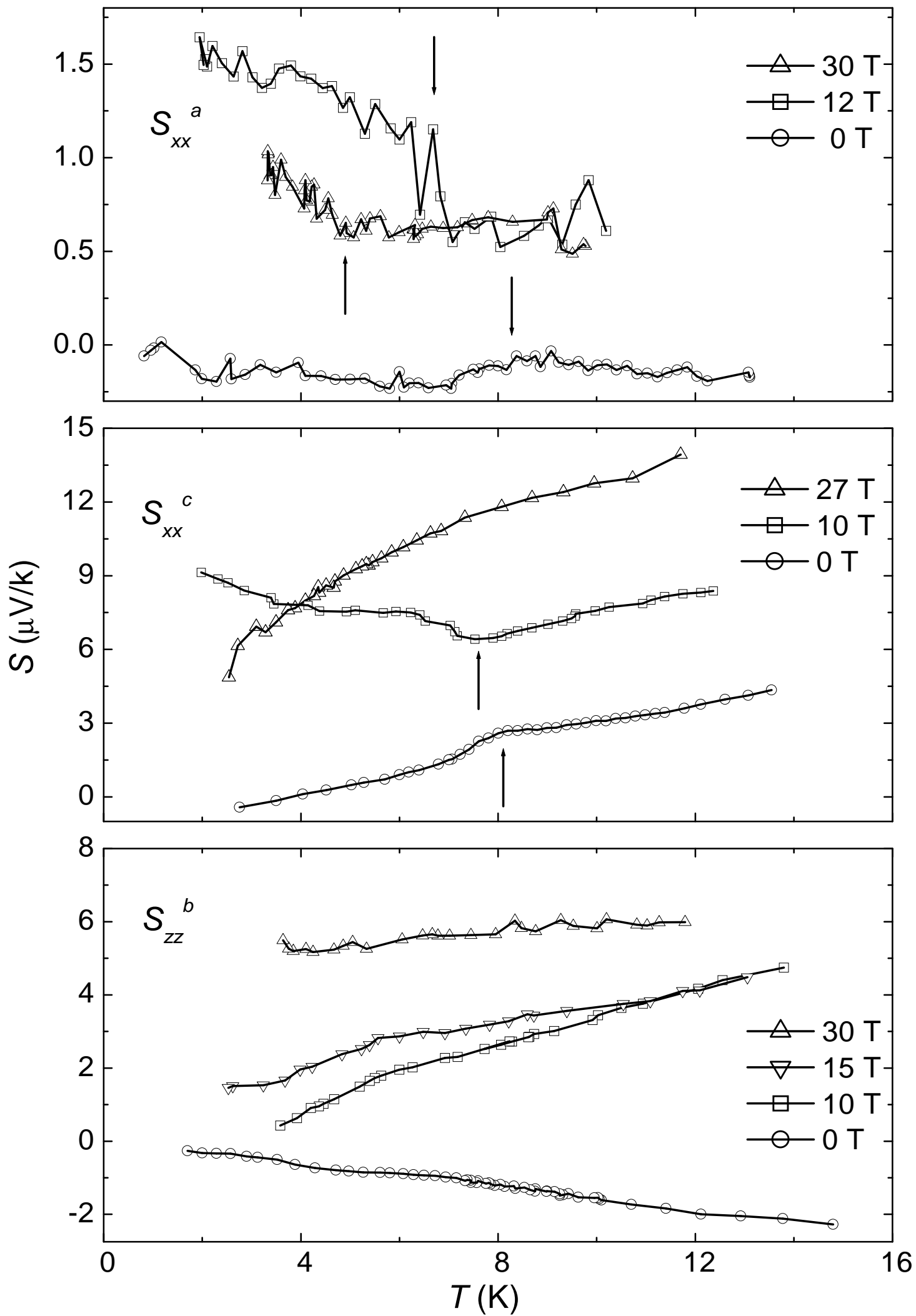
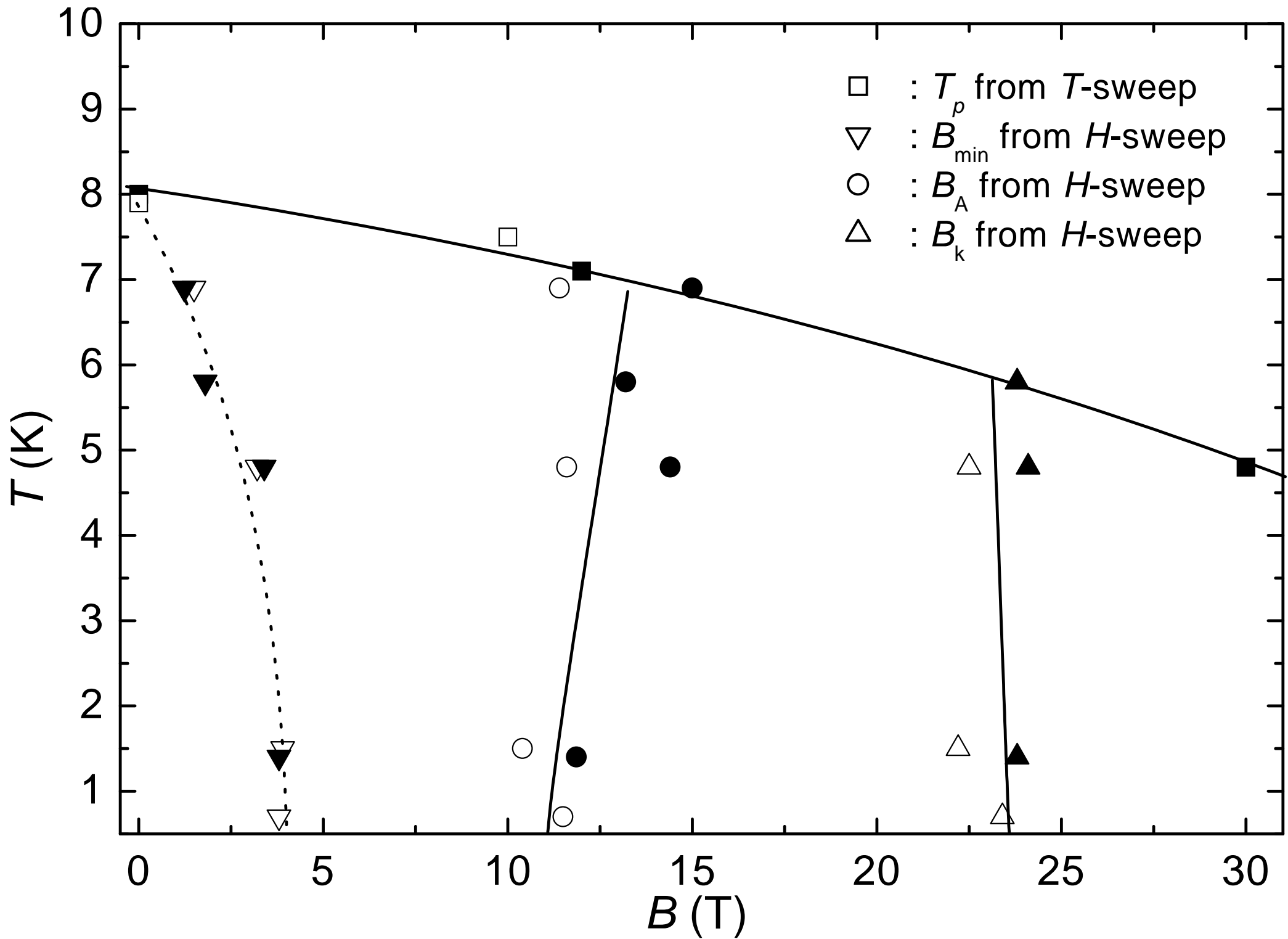
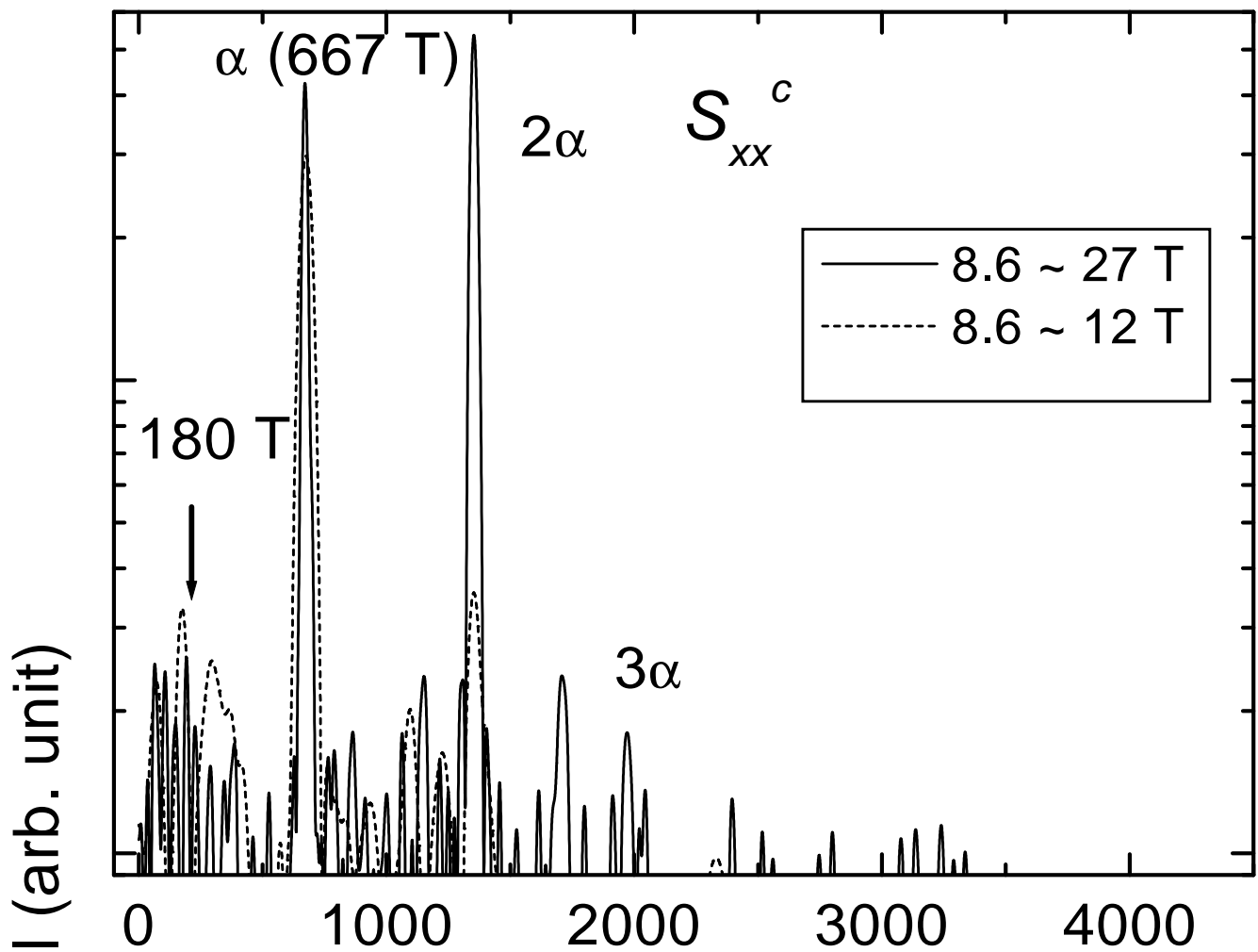
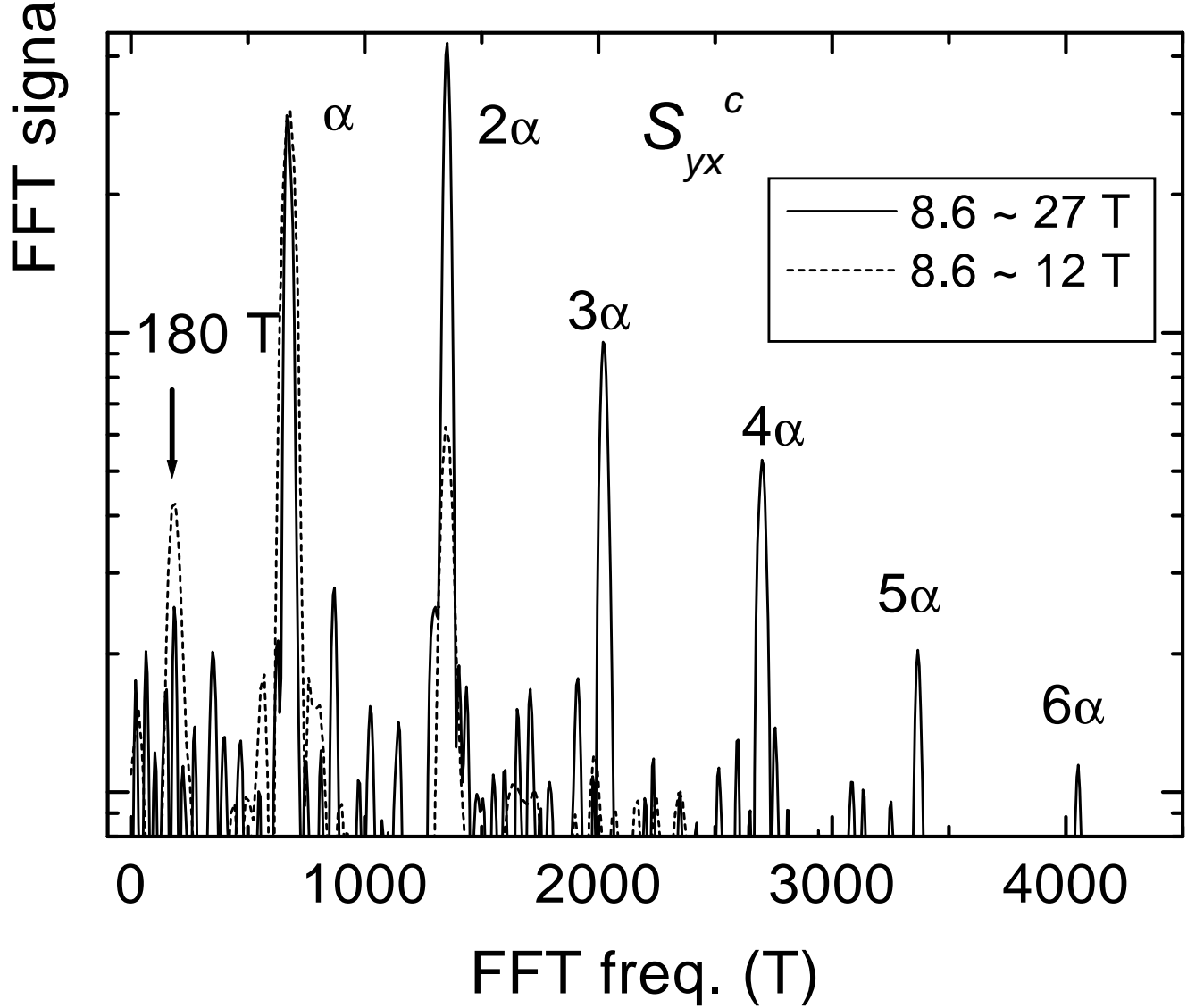
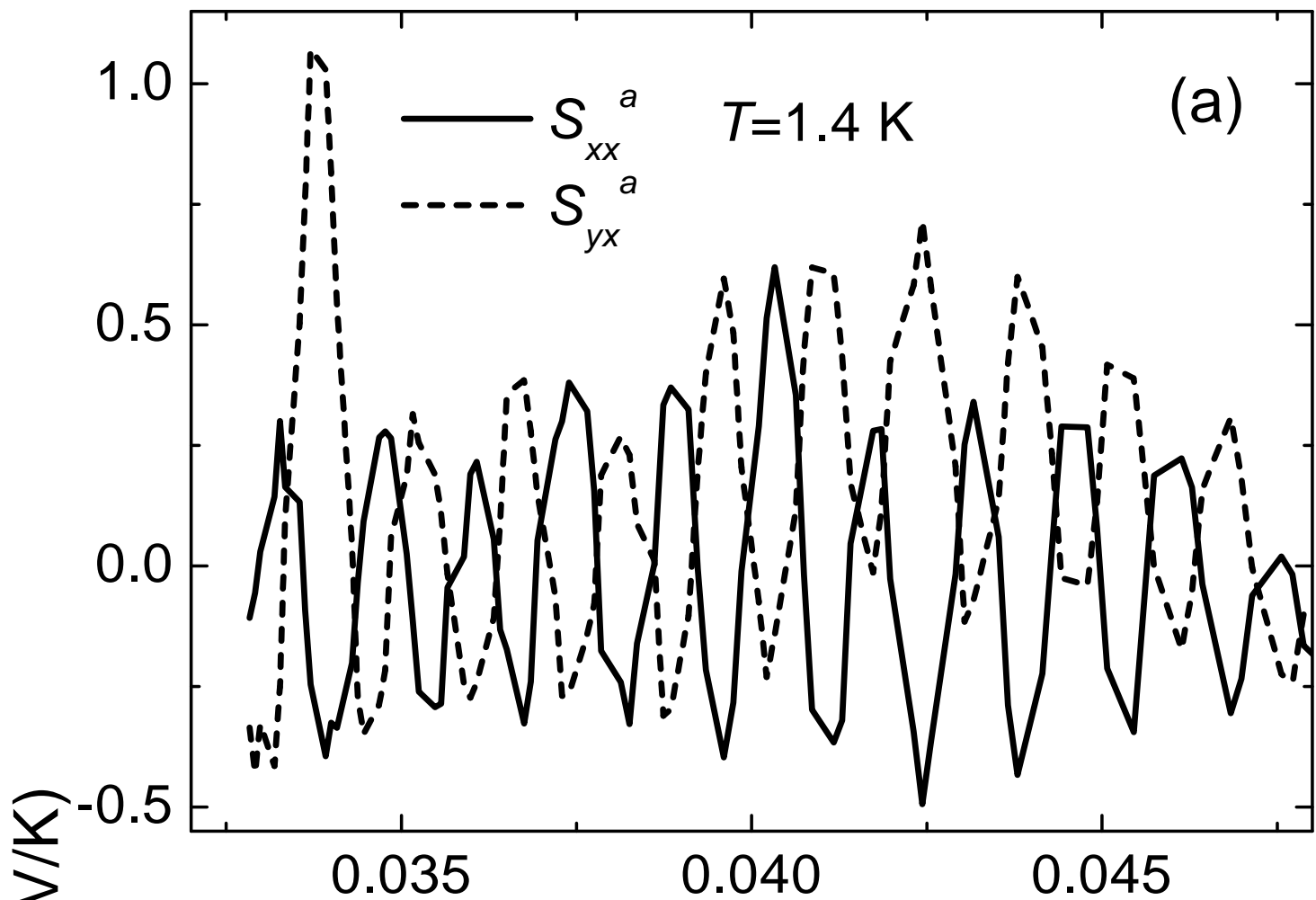
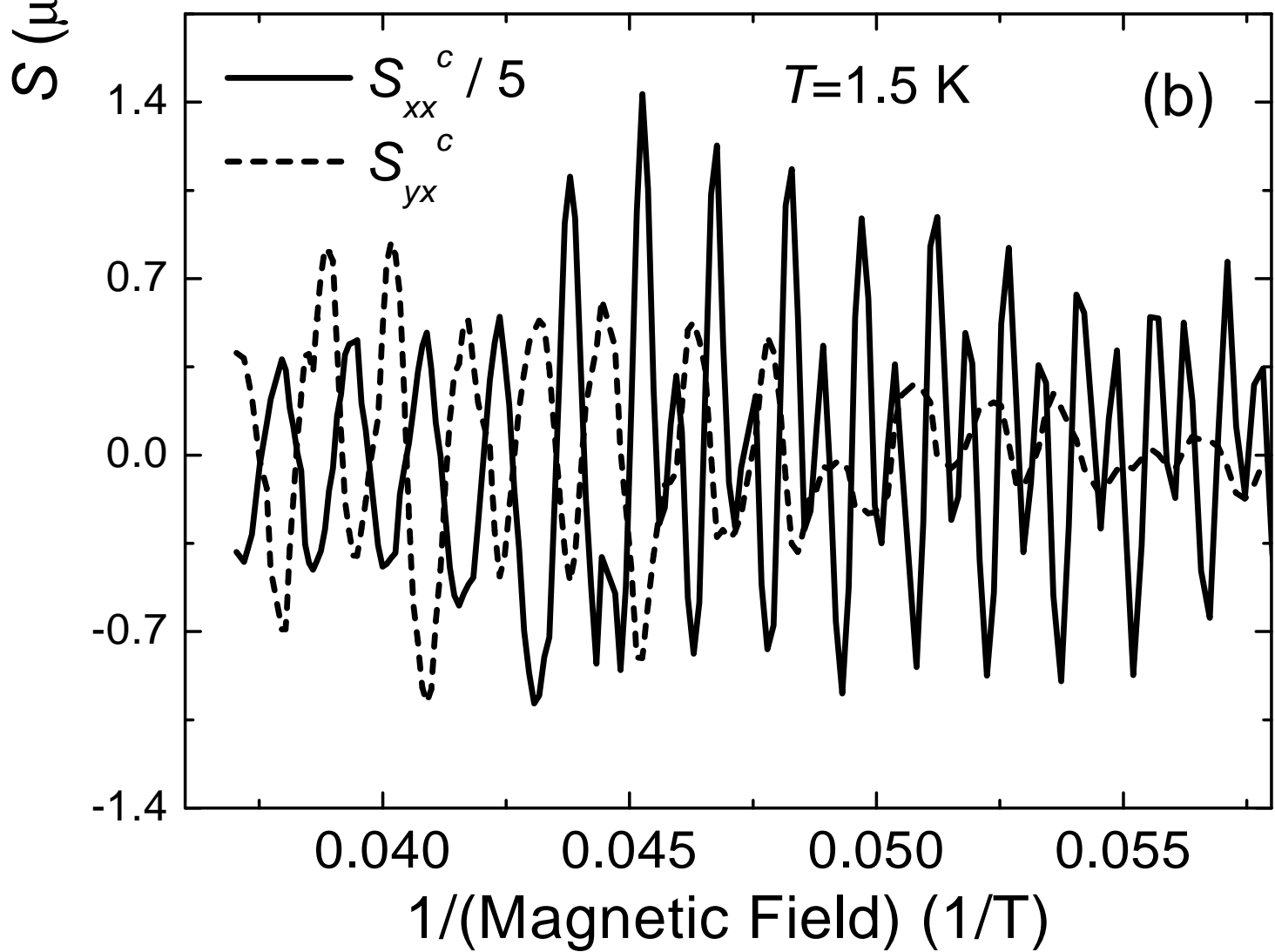
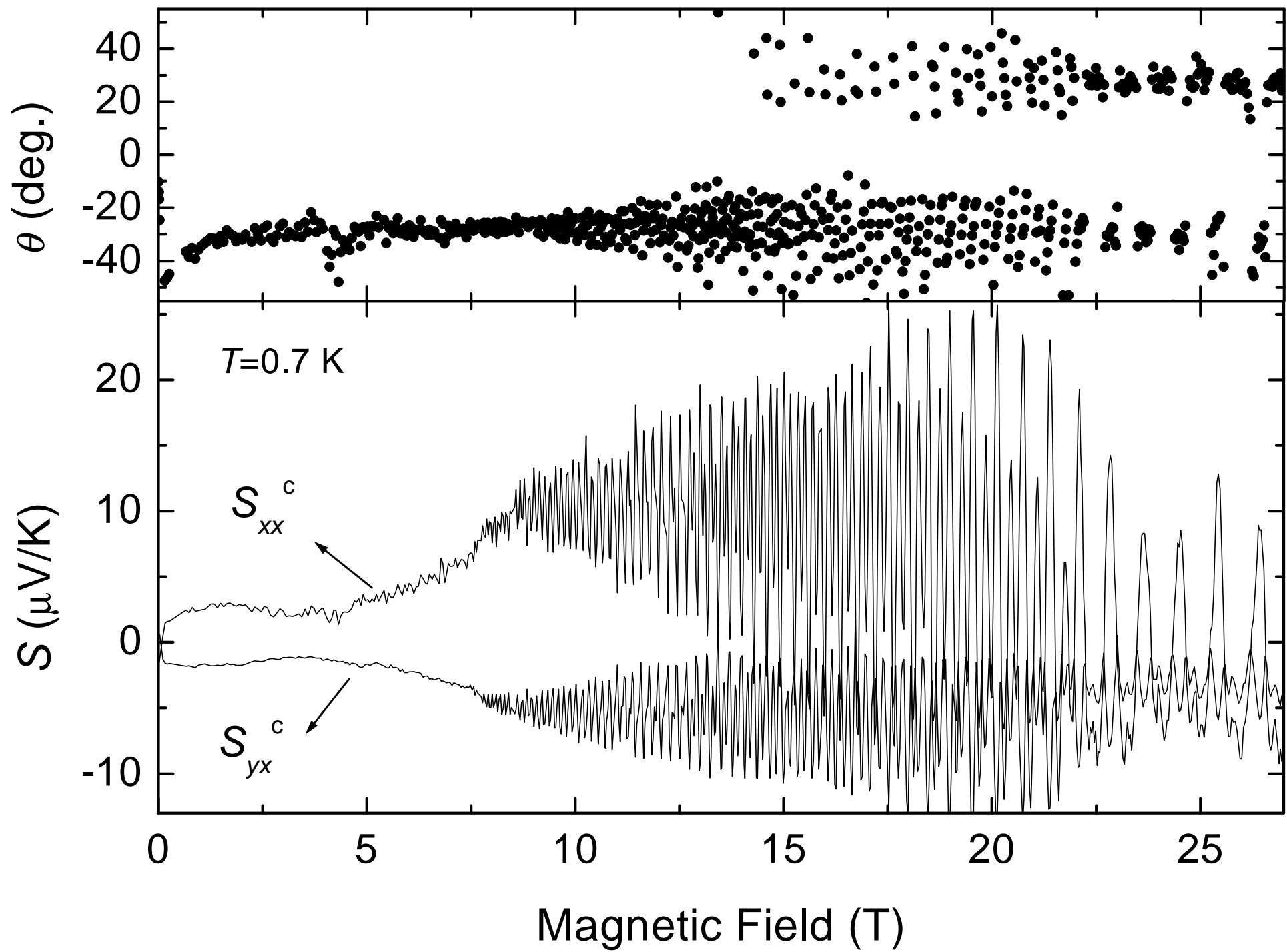
\end{document}